\documentclass{aa}  
\usepackage{graphicx}
\usepackage{txfonts}
\usepackage{color}

\newcommand{\lya}{Ly$\alpha$}
\newcommand{\hi}{H~{\small I}}

\newcommand{\mgii}{Mg~{\small II}}

\newcommand{\civ}{C~{\small IV}}
\newcommand{\heii}{He~{\small II}}
\newcommand{\ciii}{C~{\small III}]}
\newcommand{\oii}{[O~{\small II}]}

\graphicspath{{./}{figures/}}

\usepackage{hyperref}
\usepackage{natbib,twoopt}

\begin{document}

\title{Median surface-brightness profiles of Lyman-$\alpha$ haloes in the MUSE Extremely Deep Field}

\author{Yucheng Guo\inst{\ref{inst1}\thanks{e-mail: yucheng.guo@univ-lyon1.fr}}
\and Roland Bacon\inst{\ref{inst1}}
\and Lutz Wisotzki\inst{\ref{inst2}}
\and Thibault Garel\inst{\ref{inst3}}
\and J\'er\'emy Blaizot\inst{\ref{inst1}}
\and Joop Schaye\inst{\ref{inst4}}
\and Johan Richard\inst{\ref{inst1}}
\and Yohana Herrero Alonso \inst{\ref{inst2}}
\and Floriane Leclercq \inst{\ref{inst6}}
\and Leindert Boogaard \inst{\ref{inst5}}
\and Haruka Kusakabe \inst{\ref{inst3}}
\and John Pharo \inst{\ref{inst2}}
\and Eloïse Vitte \inst{\ref{inst3},\ref{inst7}}
}

\institute{Univ Lyon, Univ Lyon1, Ens de Lyon, CNRS, Centre de Recherche Astrophysique de Lyon UMR5574, F-69230, Saint-Genis-Laval, France\label{inst1}
\and Leibniz-Institut fur Astrophysik Potsdam (AIP), An der Sternwarte 16, 14482 Potsdam, Germany\label{inst2}
\and Observatoire de Geneve, Universite de Geneve, 51 Ch. des Maillettes, 1290 Versoix, Switzerland\label{inst3}
\and Leiden Observatory, Leiden University, P.O. Box 9513, 2300 RA Leiden, The Netherlands\label{inst4}
\and Department of Astronomy, University of Texas at Austin, 2515 Speedway, Austin, TX 78712, USA\label{inst6}
\and Max Planck Institute for Astronomy, K\"onigstuhl 17, 69117, Heidelberg, Germany\label{inst5}
\and ESO Vitacura, Alonso de Córdova 3107,Vitacura, Casilla 19001, Santiago de Chile, Chile \label{inst7}
}

\date{Submitted 2023}

\abstract
{
We present the median surface brightness profiles of diffuse \lya\ haloes (LAHs) around star-forming galaxies by stacking 155 spectroscopically confirmed \lya\ emitters (LAEs) at $3<z<4$ in the MUSE Extremely Deep Field (MXDF) with a median \lya\ luminosity of $\mathrm{L_{Ly\alpha} \approx 10^{41.1} erg\,s^{-1}}$.
After correcting  for a systematic surface brightness offset we identified in the data cube, we detect extended \lya\ emission out to a distance of $\approx$270~kpc. 
The median \lya\ surface-brightness profile shows a power-law decrease in the inner 20~kpc and a possible flattening trend at a greater distance.
This shape is similar for LAEs with different \lya\ luminosities, but the normalisation of the surface-brightness profile increases with luminosity.
At distances over 50~kpc, we observe a strong overlap of adjacent LAHs, and the \lya\ surface brightness is dominated by the LAHs of nearby LAEs.
We find no clear evidence of redshift evolution of the observed \lya\ profiles when comparing with samples at $4<z<5$ and $5<z<6$.
Our results are consistent with a scenario in which the inner 20~kpc of the LAH is powered by star formation in the central galaxy, while the LAH beyond a radius of 50~kpc is dominated by photons from surrounding galaxies. 
}

\maketitle
\nolinenumbers
\section{Introduction} \label{sec:intro}
Galaxies are born and bred in their gaseous haloes \citep[e.g.][]{tumlinson17}.
Within the circumgalactic medium (CGM), gas and metals can be ejected from galaxies by feedback processes or stripping, or they can be (re-)accreted to fuel star formation. 
On larger scales, the intergalactic medium (IGM) traces the cosmic web of matter that connects, forms, and fuels galaxies, and its properties may be changed by feedback processes. 
Mapping the gas distribution in the IGM and the CGM and investigating their relationship with the interstellar medium (ISM) of the galaxies are crucial for understanding the evolution of galaxies.

The hydrogen \lya\ line is a powerful tool for mapping the CGM and IGM at high redshift \citep[e.g.][]{steidel11,ouchi20,bacon21}.
Theoretical studies predict substantial amounts of neutral hydrogen in the CGM and IGM \citep[e.g.][]{fumagalli11,voort12}, and this has been observed through absorption measurements \citep[e.g.][]{lee18,newman20}. 
The \lya\ photons are produced by hydrogen recombination following ionisation or by collisional excitation following photo-heating or shock-heating.
As they propagate through the ISM, CGM, and IGM, they can be repeatedly absorbed and re-emitted by the neutral hydrogen gas.
This resonant nature of \lya\ makes it a unique tracer of neutral gas distribution, though the likelihood of it being absorbed by galactic dust is non-negligible.
In addition to its utility in mapping the hydrogen gas, the \lya\ emission line is also a highly informative probe of high-redshift star-forming galaxies \citep[e.g.][]{ouchi08,shibuya12,guo20b,bacon22}. It is intrinsically the strongest emission line in the rest-frame UV and optical spectrum \citep[e.g.][]{partridge67}.
Therefore, observing the large-scale spatial distribution of \lya\ emission can provide a map of the large-scale structure of the early Universe, including the distribution of high-redshift galaxies and the distribution of the pristine gas.

In the past decades, extended \lya\ emissions, known as \lya\ haloes (LAHs), have been observed around \lya\ emitting galaxies (LAEs) at $\mathrm{z \gtrsim 2}$ using the narrow-band (NB) technique \citep[e.g.][]{matsuda12,feldmeier13,momose14,momose16,xue17,wu20,kikuta23} and integral field unit (IFU) facilities \citep[e.g.][]{wisotzki16,leclercq17,daddi21,kusakabe22}. 
The LAHs have also been observed at low redshift in the vicinity of galaxies \citep[e.g.][]{hayes13,duval16,runnholm23}. 
The LAHs have physical scales of several tens of kiloparsecs, where most important processes regulating galaxy formation and evolution occur.
By combining deep exposures and stacking, recent studies have expanded the detection of extended \lya\ emission to hundreds of kiloparsecs \citep[e.g.][]{kakuma22,kikuchi22,maja22a,maja22b}.
Despite the substantial observational work on the spatial content of the LAHs, the physical mechanisms of the production and propagation of \lya\ photons are still under debate \citep[e.g.][]{cantalupo05,zheng11,rosdahl12,byrohl21}.

The ESO-VLT instrument Multi Unit Spectroscopic Explorer \citep[MUSE,][]{bacon10} revolutionised the observation of the CGM and IGM. This IFU facility with a large field of view is highly efficient in mapping the extended emission from the CGM, or even the IGM, using multiple tracers, e.g., \lya\ \citep[e.g.][]{wisotzki16,wisotzki18,cai17,cai19,leclercq17,leclercq20,gallego18,gallego21,claeyssens19,claeyssens22,bacon21,kusakabe22}, \civ, \heii, \ciii\ \citep[e.g.][]{borisova16,guo20},  \mgii\ \citep[e.g.][]{zabl21,leclercq22}, and \oii\ \citep[e.g.][]{johnson22}.
At $z \gtrsim 3$, MUSE observations have identified large samples of individual LAEs and allowed for the analysis of the surface-brightness profiles of the LAHs individually or by stacking.
\citet{wisotzki18} find that the LAHs on average extend from tens to hundreds of kiloparsecs, thus covering nearly all of the sky by projection.
\citet{bacon21} detect the cosmic web filaments in \lya\ emission on scales of several comoving megaparsecs.
\citet{umehata19} observe the \lya-illuminated gas filaments that connect galaxies in a protocluster.
\citet{kusakabe22} study the LAHs around UV-selected galaxies instead of objects selected from their \lya\ emission. The existence of extended \lya\ emission around non-LAEs implies a significant amount of neutral hydrogen in the CGM of the normal star-forming galaxies.
The deep MUSE observations have also allowed for the spatially resolved spectroscopic analysis of a sample of bright individual LAHs \citep[e.g.][]{claeyssens19,leclercq20}.

In this work, we measure the median \lya\ surface brightness profile of 369 spectroscopically confirmed LAEs at $3<z<6$, with our primary focus being $3<z<4$.
The current dataset is 5-14 times deeper than that of \citet{wisotzki18}.
We therefore expect to expand the detection of extended \lya\ emission out to hundreds of kiloparsecs from the LAE center. 
Previous studies have suggested that the dominant origin of the LAHs differs at different radii relative to the galaxies \citep[e.g.][]{lake15,mitchell20, byrohl21}.
By measuring the \lya\ profiles out to large radii, we expect to provide more constraints on the origin of the LAHs at different distances.

The paper is organised as follows.
We describe the definition of the LAE sample and our data reduction in Section~\ref{sec:data}.
Section~\ref{sec:results} provides the results, highlights the median \lya\ surface-brightness profile at $3<z<4$, and presents the profiles of different subsamples and datasets.
In Section~\ref{sec:discussions}, we discuss our results by comparing with previous observations and theoretical predictions, and also interpret the \lya\ profiles from the point of view of cross-correlation functions.
We summarise in Section~\ref{sec:summary}.

We adopt the standard $\mathrm{\Lambda CDM}$ cosmology with $H_0\mathrm{=70\, km\, s^{-1}\, Mpc^{-1}}$, $\mathrm{\Omega _m=0.3}$ and $\mathrm{\Omega _\lambda =0.7}$. All distances are proper, unless noted otherwise.

\section{The galaxy sample and data analysis} \label{sec:data}
This work is based on data release 2 (DR2) of the MUSE Hubble Ultra Deep Field surveys \citep{bacon22}.
The DR2 data consists of three datasets, a $\mathrm{3 \times 3}$~arcmin$^2$ mosaic of nine MUSE fields at ten-hour depth (hereafter MOSAIC), a $\mathrm{1 \times 1}$~arcmin$^2$ field at 31-hour depth (hereafter UDF-10), and the MUSE eXtremely Deep Field (MXDF), with a diameter of 1~arcmin and the deepest achieved exposure of 141 hours.
This work is mainly based on the MXDF.
It is the deepest spectroscopic survey ever performed, reaching an unresolved emission line median $\mathrm{ 1\sigma }$ surface brightness limit of $\mathrm{ <10^{-19}\,erg\,s^{-1} \, cm^{-2} \, arcsec^{-2} } $ { within a circular aperture of 1'' diameter}.
The MOSAIC dataset, which covers an area approximately nine times larger than the MXDF but has an exposure time approximately 14 times shorter, is a useful comparison dataset.
In Section~\ref{subsec:mosaic}, we describe our efforts to detect more extended emission in MOSAIC.
For further details on the data reduction of the MXDF and MOSAIC, we refer the reader to \citet{bacon22}.

\subsection{The LAE sample}
\begin{figure*}[ht!]
\centering
\includegraphics[width=0.9\textwidth]{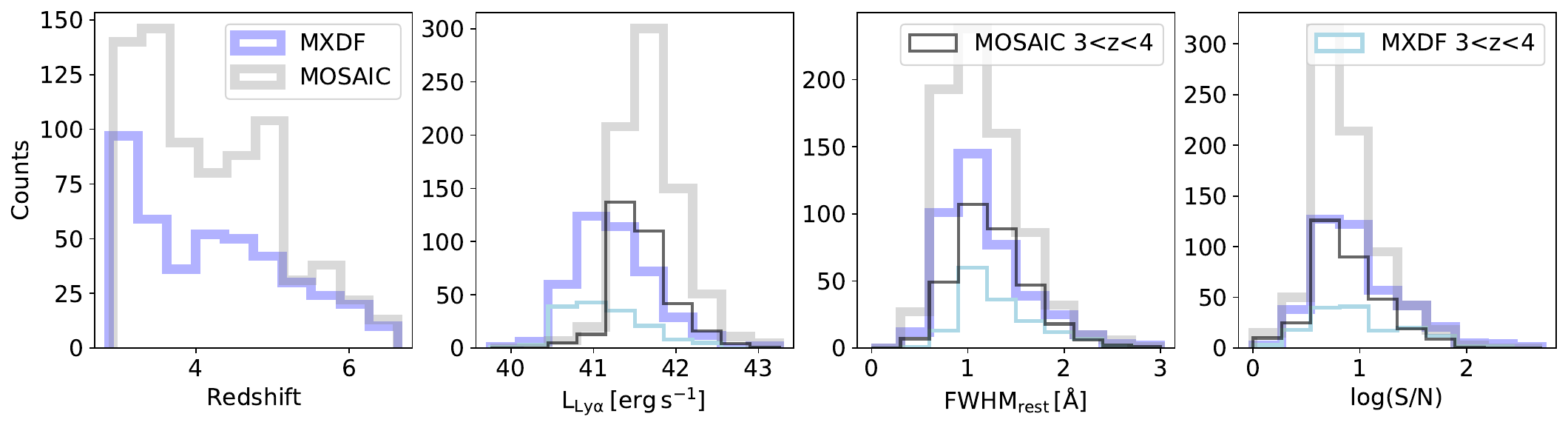}
\caption{Distribution of redshift (first panel), \lya\ luminosity (second panel), \lya\ line rest-frame FWHM (third panel), and S/N of \lya\ line (fourth panel) of the LAE sample.
The thick lines present the distributions of the whole galaxy samples in MXDF (blue lines) or MOSAIC (grey lines).
The narrow lines show the LAEs at $3<z<4$, the redshift range of most interest for this work.
\label{fig_distributions}}
\end{figure*}

\citet{bacon22} carried out a meticulous study involving multiple cross-checked procedures to achieve a high-quality and homogeneous source detection.
Their final catalogue provides the redshifts, multi-band photometries, morphological and spectral properties, as well as measurements of stellar mass and the star formation rate of all the galaxies discovered in the MOSAIC, UDF-10, and MXDF fields.
The MXDF field contains a total of 369 LAEs at $3<z<6$, while MOSAIC contains 693 detected LAEs. 
In this work, we focus on the redshift range of $3<z<4$.
In this redshift range, the efficiency of MUSE is high, the redshift dimming is low, and there are no strong OH sky lines, which enables us to acquire a highly representative LAE sample that extends to very low \lya\ luminosity. 
In the $3<z<4$ redshift range, there are 155 LAEs detected in MXDF and 329 in MOSAIC.

Figure~\ref{fig_distributions} shows the distribution of redshift, \lya\ luminosity ($\mathrm{L_{Ly\alpha}}$), rest-frame \lya\ line FWHM, and \lya\ line S/N for all LAEs and for the sample at $3<z<4$.
All these quantities are derived from \citet{bacon22}.
The median \lya\ luminosity of LAEs at $3<z<4$ in the MXDF is $\mathrm{L_{Ly\alpha, median} \approx 10^{41.1} erg\,s^{-1}}$.
The median \lya\ S/N for LAEs at $3<z<4$ in the MXDF is 7.7, with the 95th percentile of 53.4 and the 5th percentile of 2.9. 
85 of the 155 LAEs are reliably detected in HST NIR bands. Considering that all LAEs undetected in the HST broad bands have lower stellar masses, we obtain a median stellar mass of $\mathrm{M_* \approx 10^{7.6} M_\odot}$, according to the SED modelling of \citet{bacon22}.
According to \citet{herrero23}, the typical halo mass for MXDF LAEs over the whole redshift range of is $\mathrm{log(M_{h}/[h^{-1}M_{\odot}])=10.77^{+0.13}_{-0.15}}$. 
The number should be smaller for MXDF LAEs at $3<z<4$, as their typical \lya\ luminosity is lower than their parent sample (second panel of Figure~\ref{fig_distributions}).

We try to avoid contamination from the bright AGNs that can be dominant in powering the extended \lya\ emission.
Given the sky coverage of MXDF and MOSAIC (1-3~arcmin), the possibility of detecting a substantial number of AGNs is very low.
There is one type-I AGN at $z$=3.2.
There is another type-II AGN at $z$=3.06 that is cross-matched by the Chandra Chandra 7Ms catalogue \citep{luo17}. Both AGNs are found in MOSAIC \citep{bacon22} and are excluded from our analysis.
Considering that the LAEs in our sample have an \lya\ luminosity of $\mathrm{L_{Ly\alpha} < 10^{43} erg\,s^{-1}}$, the AGN contamination should be very low \citep[e.g.][]{sobral18,calhau20,zhang21}. 
According to the estimation of \citet{zhang21}, the AGN fraction at $\mathrm{L_{Ly\alpha} < 10^{43} erg\,s^{-1}}$ is $\lesssim$ 0.05. 

\subsection{Masking and continuum subtraction} \label{subsec:masking}

The exposure of the MXDF field is not uniform, with the integration time decreasing from 141 hours at the field centre to several hours at the edge of the field \citep{bacon22}.
We mask out the area with an exposure time of less than 110 hours to keep a high S/N and a more homogeneous sample throughout the selected field of view.
We also mask the wavelength slices affected by bright sky lines to avoid strong sky line residuals.
We remove the continuum by performing a spectral median filtering using a wide spectral window of 200~\AA.
This approach provides a rapid and effective way to remove continuum sources in the search for extended line emission.

We focus on detecting the extended \lya\ line at large distances from the LAEs. 
To achieve this focus, we need to remove the contamination from emission and absorption lines from foreground galaxies and from the LAEs themselves.
We mask out all the detected emission and absorption lines from all the galaxies in the continuum-subtracted data cube, with the exception of the \lya\ line.
The mask is based on the composite segmentation images provided by \citet{bacon22}.
The segmentation images are provided by SExtractor \citep{bertin96}, with an S/N limit of 2.

We note that if there is more than one LAE in a NB, we keep all the objects.
In Section~\ref{subsec:neighbors}, we demonstrate that as the exposure becomes deeper and the distance increases, neighbouring LAEs inevitably appear within the field of view. Therefore, in Section~\ref{subsec:neighbors}, we also present an additional measurement after removing the neighbouring LAEs.

\begin{figure}[t!]
\centering
\includegraphics[width=0.46\textwidth]{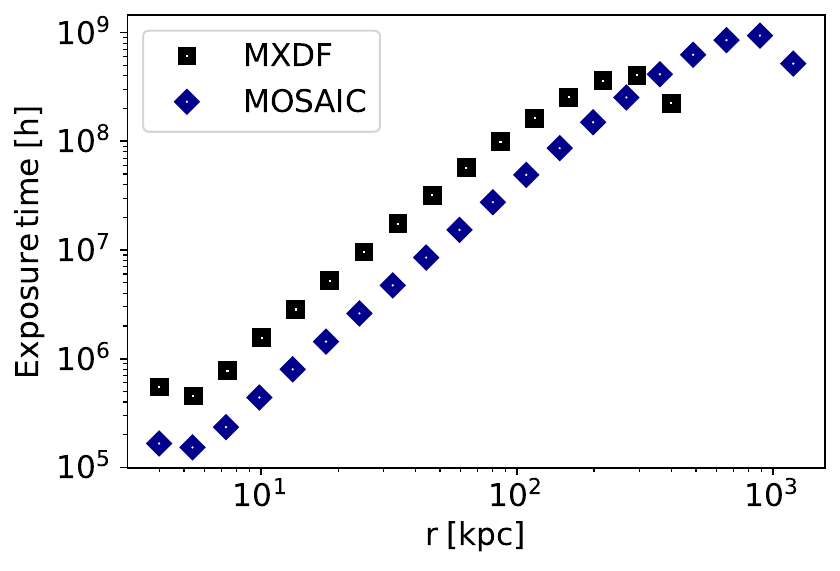}
\caption{
Stacked exposure time in each radial bin for LAEs at $3<z<4$. The black and blue symbols show the MXDF and MOSAIC sample, respectively.
\label{fig_expt}}
\end{figure}
\subsection{Extraction of \lya\ surface brightness} \label{subsec:extraction}

We construct the pseudo-NB image for each LAE based on the MUSE data cube. 
Each NB is centred on the peak wavelength of the \lya\ line. 
In the case of \lya\ lines with two peaks, we centre the NB on the red peak.
{ 
The spectral bandwidth of the \lya\ NB is approximately 15.0~\AA\ in the observational frame (924.4~km/s) at $z\approx 3$, and 34.2~\AA\ (1206.4~km/s) at $z\approx 6$. The chosen bandwidth corresponds to a fixed comoving distance of 12~cMpc.
}
This bandwidth is chosen to include most of the \lya\ flux, providing a reasonable tolerance for the variety of \lya\ line widths (Figure~\ref{fig_distributions}), uncertainty in the measurement of the \lya\ peak wavelength, and the spatial variation of the \lya\ line at large radii (Guo et al. 2023 in preparation).

In each NB image, we measure the surface brightness of the \lya\ emission in radial bins.
We average the flux inside successive concentric annuli centred on each LAE, with the central radii of annuli ranging from 10 to 270 kpc.
The final annulus includes signal as distant as 470~kpc.
We ignore the possible spatial offset of the \lya\ centroid and the UV centroid, as our point of interest is the extended emission at large radius. 
For bright $2.9<z<6.7$ LAEs that can be detected both in UV and \lya, the median spatial offset between UV and \lya\ emission is reported to be 0.58$\pm$0.14~kpc \citep{claeyssens22}.
Therefore, most of the possible offsets between the UV and \lya\ should reside in the first radial bin ($\approx$1.4'') and have a negligible impact on the detection at large distances.

Finally, we stack all individual radial profiles to obtain the median \lya\ surface brightness profile of the LAH.
In order to retain physical units, we do not normalize the surface brightness profile of each LAE.
As is shown in Section~\ref{sec:results}, we present the results for different redshift intervals and different $\mathrm{L_{Ly\alpha}}$ bins.

We note that in this method we make use of the full potential of the MXDF's field of view. 
As we extend to the furthest radial bin, it approaches the edge of the field. 
Consequently, the limited area of the field introduces uncertainties associated with reduced exposure time.
In Figure~\ref{fig_expt}, we sum up the exposure time of all pixels within each radial bin and plot it against radial distance. 
At larger distances ($\gtrsim$300~kpc for the MXDF, and $\gtrsim$800~kpc for the MOSAIC), the stacked exposure times indeed start to decrease due to field limitations.
However, for the distances of primary interest in this study (within $\approx$100~kpc), this effect is negligible. 

Our methods for signal extraction and stacking are not exactly the same as used in previous works of the same instrument \citep[e.g.][]{wisotzki18}. 
We first extract the individual radial profiles and then median-stack those profiles, while \citet{wisotzki18} measured the radial profile from the median-stacked NB image.
In this work, we defined the annuli in physical distance, whereas \citet{wisotzki18} stacked pseudo-NBs of LAEs at different redshifts in the MUSE observed frame. 
Our method has the advantage that we include the average of any anisotropic (or even filamentary) emission at large distances.
The method of \citet{wisotzki18} has the advantage that the median-stacking of the images screens out the signals from the neighbours, while those contributions are inevitable in our profiles. In Section~\ref{subsec:neighbors}, we further quantify the contribution of the neighbours detected in the field.

\subsection{Estimation of the surface brightness systematic offset} \label{subsec:systematic}

Although \citet{bacon22} provided data cubes that underwent very high-quality data reduction and sky subtraction, a residual background may still exist, including the sky residuals or other unknown systematic effects.
In this work, we refer to this residual background as systematic surface brightness offset.
At low surface-brightness levels, the uncertainties introduced by the systematic offset may not be negligible. Here, we estimate this offset using two independent methods.

\begin{itemize}

\item
$Random \; Wavelength$. 
First, we randomly selected positions in the data cube at the same sky coordinate as a real LAE, but at random wavelength layers. 
These positions are adjacent to the real \lya\ line within a wavelength range of 5~\AA\ to 50~\AA\ (rest-frame).
We extracted the surface-brightness profiles from these randomly selected positions using the same procedures employed as with the real LAEs. 
As the data cube is continuum-subtracted, the surface-brightness profiles should be zero, on average. If the stacked profile is not zero, we can infer that a systematic offset is present.
We generated 100 of these random measurements for each object.
We then estimate the potential systematic offset by taking the median of these random surface brightness profiles.

\item
$Random \; Position$. 
We also tried another method by picking up random positions in the pseudo-NB image.
For each real LAE, we took 100 such measurements at random positions that are at least 8'' away, but within the same image.
We then extracted the surface-brightness profiles from these random positions. 
The median of these random profiles is then taken as another measurement of the systematic offset.

\end{itemize}

The systematic surface-brightness offsets are finally subtracted from the surface-brightness profile of each individual LAE before we stack all the profiles.
As illustrated in the lower panels of Figure~\ref{fig_profile_all_unmask}, the two measurements of systematic surface-brightness offsets exhibit good agreement with each other. The median systematic offset is very small ($\mathrm{ \approx 1.67 \times\ 10^{-21}\,erg\,s^{-1} \, cm^{-2} \, arcsec^{-2} }$).
To estimate the stacking uncertainty, we employ a bootstrap algorithm, randomly resampling 70\%\ of the sample for 10,000 iterations.

\section{Results} \label{sec:results}
\subsection{The median \lya\ surface brightness profiles}
\label{subsec:results31}
\begin{figure*}[ht!]
\centering
\includegraphics[width=0.9\textwidth]{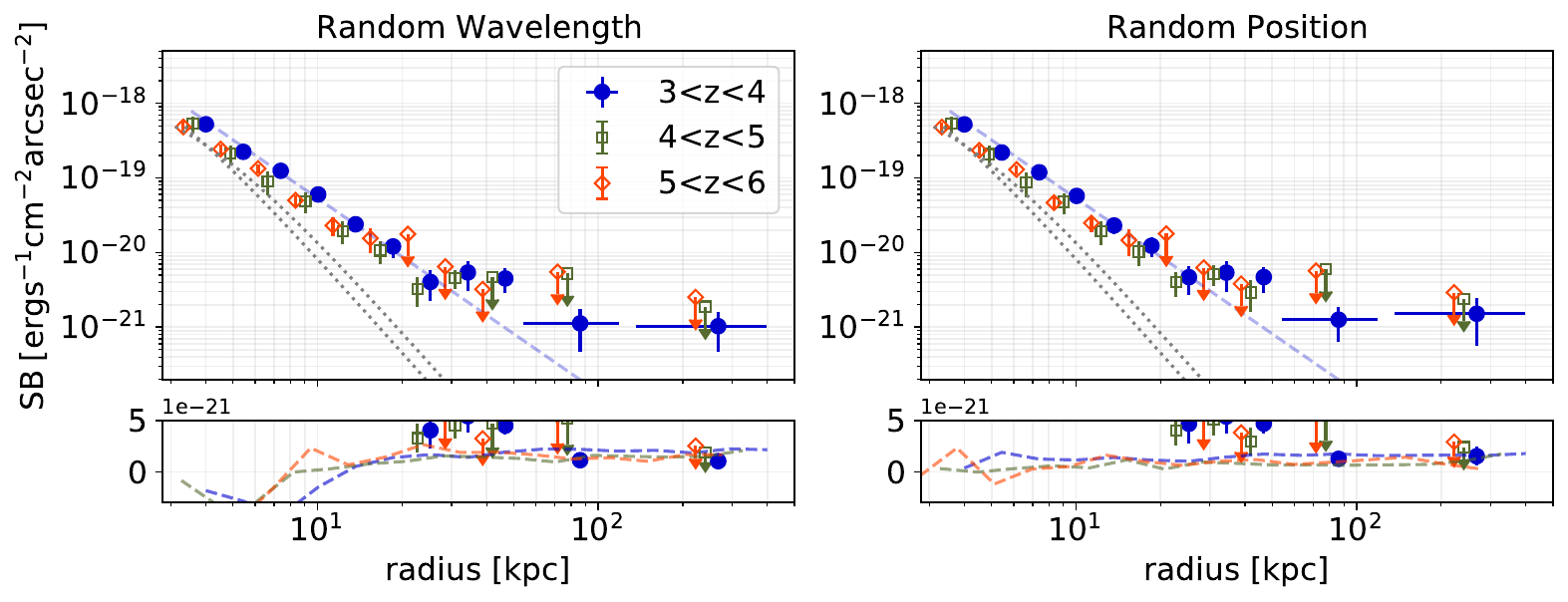}
\caption{Median \lya\ surface-brightness profiles of our MXDF LAE sample after subtracting the systematic surface brightness offset. 
The left and right columns present the two methods for measuring the systematic offset (Section~\ref{subsec:systematic}) by inserting fake objects at the same sky position as the real LAEs but in different wavelength layers (left column) and by inserting fake objects in the same wavelength layers but at different sky positions (right column). 
The blue dashed lines show the power-law fit.
The small panels at the bottom show the systematic offset on a linear $y$ scale. 
The final radial bin is centred at 270~kpc, but it includes signal out to $\approx$470~kpc.
Different colours show different redshift ranges. 
All the symbols are slightly shifted horizontally for better visualisation.
We plot the PSFs in the MXDF at 3000\AA\ and 9000\AA\ for comparison (dotted lines). 
\label{fig_profile_all_unmask}}
\end{figure*}

Figure~\ref{fig_profile_all_unmask} presents the median \lya\ surface-brightness profiles around LAEs in different redshift ranges.
The left and right panels show the two measurements of the systematic surface-brightness offset. 
The PSFs of the MXDF at 5000\AA\ and 9000\AA\ are also plotted for comparison. 
For all redshift ranges, the median \lya\ surface brightness profiles are clearly more extended than the PSF.

In the left and right columns of Figure~\ref{fig_profile_all_unmask}, we compare the surface brightness profiles correcting from the systematic offset estimated using the two independent methods presented in Section~\ref{subsec:systematic}.
These two independent measurements give consistent results. 
In the following sections we only show the results of the $random- \; wavelength$ method.
However, we verified that the results of the $random \; position$ or the average of these two methods does not change our conclusions.

As shown by the blue dots in Figure~\ref{fig_profile_all_unmask}, at $3<z<4$, we detect the extended \lya\ emission out to hundreds of kiloparsecs.
Within approximately 20~kpc, the \lya\ surface brightness exhibits a gradual decrease.
If we consider solely the scattering of \lya\ photons from the central galaxy to the CGM,  a power-law surface-brightness profile with a slope of -2.4 has been predicted \citep{kakiichi18}.
We apply a power-law model to our stacked profile, specifically targeting the inner $\approx$20~kpc while also accounting for the PSF effect.
The resulting power-law index, $\mathrm{-2.47 \pm 0.11,}$ agrees with the theoretical prediction, suggesting a notable influence of the scattering effect within this distance.

At radii larger than 20~kpc, a noticeable flattening trend emerges, which is more plausible within the range of 20 to 50~kpc.
Within this radial span, the \lya\ surface brightness stays at a level of $\mathrm{ 4.73 \pm 1.95 \times\ 10^{-21}\,erg\,s^{-1} \, cm^{-2} \, arcsec^{-2} }$.
The detections within the 20--50~kpc range have a significance slightly above 2$\sigma,$  which means we cannot entirely rule out variations attributed to errors. 
Nevertheless, as demonstrated in Section~\ref{subsec:mosaic}, this change of slope in the \lya\ surface-brightness profile at large radii is consistently observed across different data samples and facilities \citep[e.g.][]{wisotzki18,maja22a}.

The furthest radial bin includes signals as far out as 470~kpc, with its centre located at approximately 270 kpc. 
Within the 50--270~kpc range, we detect a tentative signal, with the combined surface brightness $\mathrm{ 1.20 \pm 0.49 \times\ 10^{-21}\,erg\,s^{-1} \, cm^{-2} \, arcsec^{-2} }$.
Despite the presence of errors, a comparison with the 20--50~kpc range reveals a break in the \lya\ surface brightness at 50--270 kpc, with a drop occurring after approximately 50 kpc.

At higher redshift, detection is limited to smaller distances. We achieve a robust detection within approximately 30~kpc and 15~kpc for LAEs at $4<z<5$ and $5<z<6$, respectively.
Comparing the three redshift intervals, we do not detect any obvious evolution of the observed \lya\ surface brightness profiles. This is in agreement with previous works \citep[e.g.][]{wisotzki18}.
When we compare the \lya\ surface-brightness profiles at different redshifts, we cannot ignore the observational effects.
Due to the stronger cosmological surface brightness dimming at higher redshift, lower efficiency of MUSE at the red end and the stronger sky background contamination, the high-redshift LAEs in our sample are more luminous than those at $3<z<4$ (see the second panel of Figure~\ref{fig_distributions}).
As we show in Section~\ref{subsec:subsamples}, there is a strong correlation between the \lya\ luminosity of the galaxy and the brightness of the LAH.
Therefore, observational incompleteness may complicate the comparison of the \lya\ surface-brightness profiles at different redshifts.

Because we detect the extended \lya\ haloes out to hundreds of kiloparsecs, the neighbouring galaxies within the field of view, if they exist, may contribute to the surface-brightness profile and are more likely to do so with increasing distance.
In Section~\ref{subsec:neighbors}, we mask all nearby LAEs to provide a `cleaner' \lya\ surface-brightness profile.
However, since these neighbours are quite common, and we are not able to distinguish between two adjacent neighbouring LAHs, we take Figure~\ref{fig_profile_all_unmask} as the more valuable result; it contains all the averaged environmental information at large distances.

\subsection{Influence of the nearby galaxies} \label{subsec:neighbors}

\begin{figure}[ht!]
\centering
\includegraphics[height=17cm]{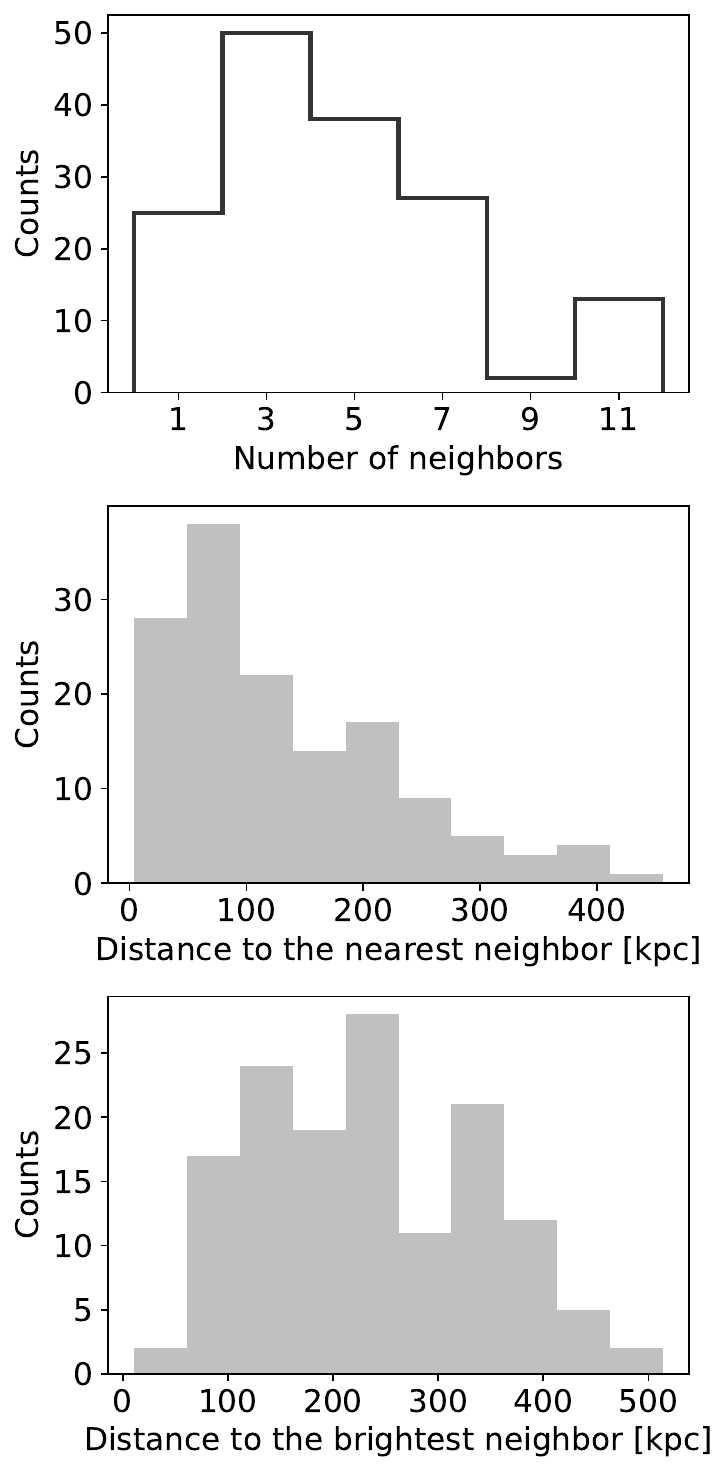}
\caption{
Distribution in upper panel presents number of neighbours detected within the MXDF field for each MXDF LAE at $3<z<4$.
The middle panel displays the projected distance to the nearest neighbour detected within the MXDF field of view.
The lower panel displays the projected distance to the brightest neighbour detected within the MXDF field of view.
All three panels are based on the MXDF galaxy catalogue \citep{bacon22}.
\label{fig_LAE_neighbors}}
\end{figure}

\begin{figure}
\centering
\includegraphics[width=0.48\textwidth]{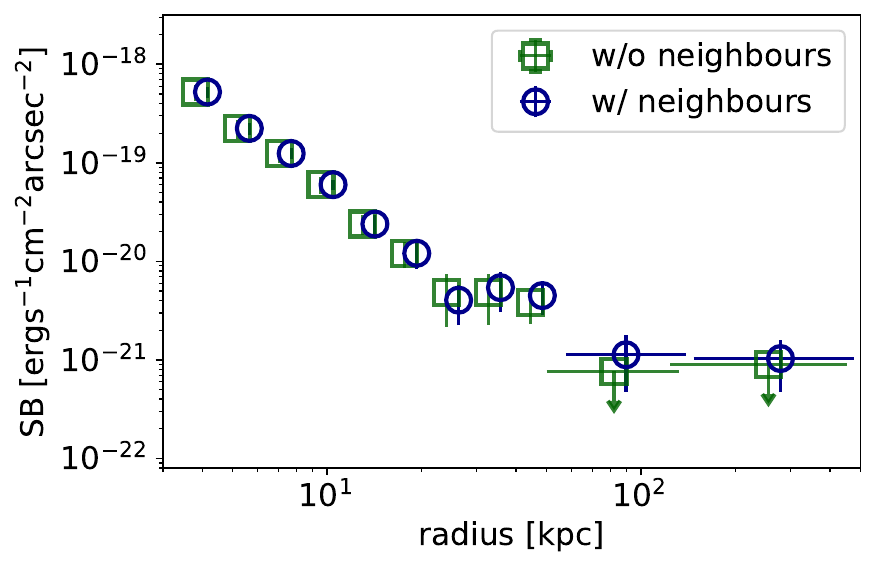}
\caption{
 \lya\ surface-brightness profiles after masking the neighbouring LAEs (green symbols) at $3<z<4$. For comparison, we also plot the profiles before masking the neighbouring LAEs (blue symbols).
The symbols are slightly shifted horizontally for better visualisation.
\label{fig_profile_compare_mask}}
\end{figure}

As we mention in Section~\ref{subsec:masking}, since we detect the extended \lya\ emission at hundreds of kiloparsecs, the influence of nearby LAEs (or the nearby LAHs) is inevitable. 
This is demonstrated by our statistics in Figure~\ref{fig_LAE_neighbors}.
In the upper panel of Figure~\ref{fig_LAE_neighbors}, we calculate the number of neighbours for each LAE at $3<z<4$ in the MXDF.
Here, the neighbour of an LAE is defined as any LAE at the field of view of the MXDF and its \lya\ peak wavelength $\pm$ 0.5$\times$FWHM at the wavelength range of the same NB.
The LAE and its neighbour(s) are within a projected distance of 1', but not necessarily physically or gravitationally related.
We find that out of 155 MXDF LAEs at $3<z<4$, only 14 LAEs have no detected neighbours within the field of view of the MXDF. 
Although it is not shown in Figure~\ref{fig_LAE_neighbors}, if we also consider the LAEs in the MOSAIC, which includes the brighter neighbours at larger distances, all the 155 MXDF LAEs have at least one neighbour. The highest number of neighbours for an LAE can exceed 30.

This statistic in the top panel of Figure~\ref{fig_LAE_neighbors} is highly observationally biased, because the sample incompleteness varies with redshift and because neighbours may exist outside the MXDF sky area.
According to the analyses of \citep{bacon21}, 70 out of 155 MXDF LAEs at $3<z<4$ are located in 14 galaxy groups. Despite these group galaxies, there is still a very high probability that other blank-field LAEs have at least one neighbour within a few hundred kpc according to our statistics. 
The diffuse \lya\ emission from these neighbours thus alters the \lya\ surface-brightness profile at the relevant distance.

In the middle panel of Figure~\ref{fig_LAE_neighbors}, we present the distance of an MXDF LAE to its nearest neighbour detected in the MXDF sky area.
In the lower panel of Figure~\ref{fig_LAE_neighbors}, we present the distance of a MXDF LAE to the neighbour with the highest \lya\ luminosity detected in the MXDF sky area.
We note that the calculation is limited by the shape of the deep field as well as the depth of the observation; there may be closer neighbours that are outside the MXDF sky area and/or neighbours too faint to be detected.
Despite these factors, on the scale of hundreds of kiloparsecs we see the overlapping of the extended \lya\ emission of adjacent LAEs.
Compared to previous work, this LAH-overlapping problem is more evident for us, because our extremely deep observations identify lots of faint LAEs that could not be detected in previous observations.
In fact, the previous work, restricted by observational depth or spatial resolution, cannot remove the effect of the neighbouring LAEs and achieve a `cleaner' radial profile.
In other words, the previous observations with shorter integration time or lower spatial resolution already included lots of faint neighbours in their \lya\ surface-brightness profiles.

We provide another measurement of the \lya\ surface-brightness profile by excluding the nearby LAEs.
In each NB, except the target LAE, we mask all the nearby galaxies based on the segmentation images given by \citet{bacon22}. Then, we compute the median profiles following the same procedures as in Section~\ref{subsec:extraction}.
This operation only removes the bright area of the nearby LAHs, because we are not able to decompose each galaxy from its LAH.
The final surface-brightness profile may still contain the diffuse content of the nearby LAHs.

The median \lya\ surface brightness profile after masking the neighbouring LAEs is shown by the green dots in Figure~\ref{fig_profile_compare_mask}.
For comparison, we also show the \lya\ surface-brightness profile before masking the neighbours (blue dots in Figure~\ref{fig_profile_compare_mask}; same as Figure~\ref{fig_profile_all_unmask}).
Within the inner 50~kpc, the two profiles are very similar, though there are inconspicuous differences at 20-50~kpc. 
Beyond 50~kpc, we do not obtain a robust detection after masking the neighbours.
The upper limits in the last two radial bins further show the break of the \lya\ surface-brightness profile at $\approx$ 50~kpc. 
Limited by the radial sampling and the S/N, we are not able to provide better constraints on this break, in particular its exact distance. However, the downward trend of the profile is quite robust.
After stacking the outer two radial bins, we provide a two-sigma upper limit of the \lya\ surface brightness at 50-270~kpc of $\mathrm{ 0.93 \times\ 10^{-21}\,erg\,s^{-1} \, cm^{-2} \, arcsec^{-2} }$, which is more than half an order of magnitude lower than the surface brightness at 20-50~kpc.
 
The potential flattening of the \lya\ surface brightness profile beyond 20~kpc is intriguing. 
This observation suggests a change in the dominant mechanism(s) responsible for producing \lya\ photons at varying distances. 
Moreover, the possible break at around 50~kpc likely indicates that the mechanisms governing the production and propagation of \lya\ photons at smaller radii become less influential at larger distances.

{ Despite the very low surface brightness of the extended emission}, the neighbouring LAEs at distances larger than 50~kpc contribute significantly to the total observed \lya\ flux.
By integrating the surface-brightness profiles over radial distance, we are able to estimate the total \lya\ photon budget around a typical LAE at $3<z<4$.
Within a distance of 270~kpc, the total \lya\ flux is about $\mathrm{ 9.07 \times\ 10^{-18}\,erg\,s^{-1} \, cm^{-2} }$, corresponding to a total \lya\ luminosity of about $\mathrm{ 2.92 \times\ 10^{41}\,erg\,s^{-1}}$.
Of the photon budget, approximately 73\%\ of the flux arises from distances larger than 50~kpc, while approximately 17\%\ and 10\%\ of the flux come from $\mathrm{r \lesssim 20~kpc}$ and $\mathrm{20 \lesssim r \lesssim 50~kpc}$, respectively.

The MXDF field is also covered by deep HST ACS F606W and F775W observations. The median magnitudes of these two filters are 29.84~mag and 29.86~mag, respectively. We average these two values to estimate the typical UV luminosity (approximately $\mathrm{ 3.8 \times\ 10^{38}\,erg\,s^{-1} \, \AA^{-1} }$). Combining the total \lya\ luminosity within 270~kpc, we give an estimate of the rest-frame \lya\ EW of 760.2~\AA. 
This number is significantly larger than the theoretical upper limit \citep[$\approx$200~\AA; e.g.][]{charlot93}, which means that the \lya\ photon budget within this distance cannot be solely explained by a `photoionisation + recombination' scenario of a central galaxy, except for certain extreme circumstances, including very low ISM metallicity \citep[e.g.][]{maseda23}.
{However, if we consider that an LAE typically has 2-4 neighbours and the continuum of these neighbours is negligible, then the average \lya\ EW of the target LAE and its neighbours is about 150 -- 250 \AA, which is roughly in line with the expectations of stellar population synthesis \citep{charlot93}.
If we only consider the \lya\ photons from the inner 50~kpc, the resulting \lya\ EW is about 205.3~\AA.}
The assumption that the inner 20~kpc is dominated by the central galaxy yields a \lya\ EW of 129.2~\AA, in agreement with previous MUSE analyses \citep{hashimoto17}.
In Section~\ref{subsec:physical_origin}, we further discuss the possible dominant mechanisms at different distances.

\subsection{LAHs of bright and faint LAEs}
\label{subsec:subsamples}

\begin{figure}
\centering
\includegraphics[width=0.48\textwidth]{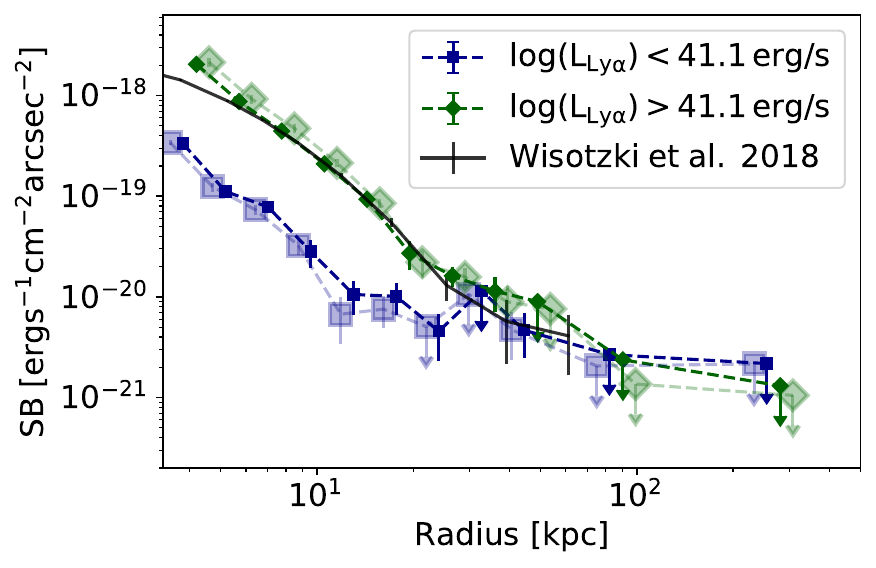}
\caption{\lya\ surface-brightness profiles at $3<z<4$. We split the LAE sample into two subsamples based on $\mathrm{L_{Ly\alpha}}$. 
The profiles before masking the neighbouring LAEs are shown by small symbols.
The profiles after masking are shown by large, light-coloured symbols. 
For comparison, we also plot the result of \citet{wisotzki18}. 
The symbols are slightly shifted horizontally for better visualisation.
\label{fig_bins}}
\end{figure}

To study the possible correlations between \lya\ surface-brightness profiles and LAE properties, we divide the LAE sample at $3<z<4$ into two subsamples based on the median \lya\ luminosity $\mathrm{L_{Ly\alpha}}$. 
The corresponding \lya\ surface brightness profiles are shown in
Figure~\ref{fig_bins}. The green and blue dots present the high- and low-$\mathrm{L_{Ly\alpha}}$ subsamples, respectively.
For each subsample, we show the profiles before and after masking the neighbouring galaxies.

For all subsamples, the \lya\ surface-brightness profiles decrease within approximately 20~kpc. 
The \lya\ profiles of high- and low-$\mathrm{L_{Ly\alpha}}$ LAEs have similar slopes, but the normalisation is $\approx$1-dex fainter for the low-$\mathrm{L_{Ly\alpha}}$ subsample.
The positive correlation between the \lya\ luminosity of the central LAE and the normalisation of the \lya\ surface-brightness profile at $\lesssim$20~kpc suggests that the production of \lya\ photons in the central galaxies plays an important role in the boosting of LAHs at this distance.

At larger radii ($\gtrsim $20~kpc), the two radial profiles show evidence of flattening, despite the large errors (or upper limits).
The flattening trend appears for both subsamples, but the change occurs at a shorter distance for the low-$\mathrm{L_{Ly\alpha}}$ subsample.

As we discuss in Section~\ref{subsec:neighbors}, nearby LAEs are contributors to the LAHs at distances larger than 50~kpc.
The trend is not very clear for the two subsamples because the S/N is low. 
Still, the values (or upper limits) measured before masking the neighbouring galaxies are slightly higher than those after masking them.
Nonetheless, the contribution from neighbouring galaxies may not be the key reason for the flattening of the profiles at 20-50~kpc, which is already demonstrated in Section~\ref{subsec:neighbors} and Figure~\ref{fig_profile_compare_mask}.

The \lya\ surface-brightness profile of higher-$\mathrm{L_{Ly\alpha}}$ LAEs agrees with \citet{wisotzki18}. 
The sample of \citet{wisotzki18} is built from brighter LAEs in the UDF-10 and MOSAIC fields.
The good agreement between \citet{wisotzki18} and our high-$\mathrm{L_{Ly\alpha}}$ profile further demonstrates the robustness of our measurement.
Similar to our work here, \citet{wisotzki18} split their LAE sample by $\mathrm{L_{Ly\alpha}}$. They also reach a similar conclusion that the radial profiles follow the same trend but with different normalisations of the surface brightness.

\subsection{Comparison of different datasets}
\label{subsec:mosaic}

\begin{figure}
\centering
\includegraphics[width=0.48\textwidth]{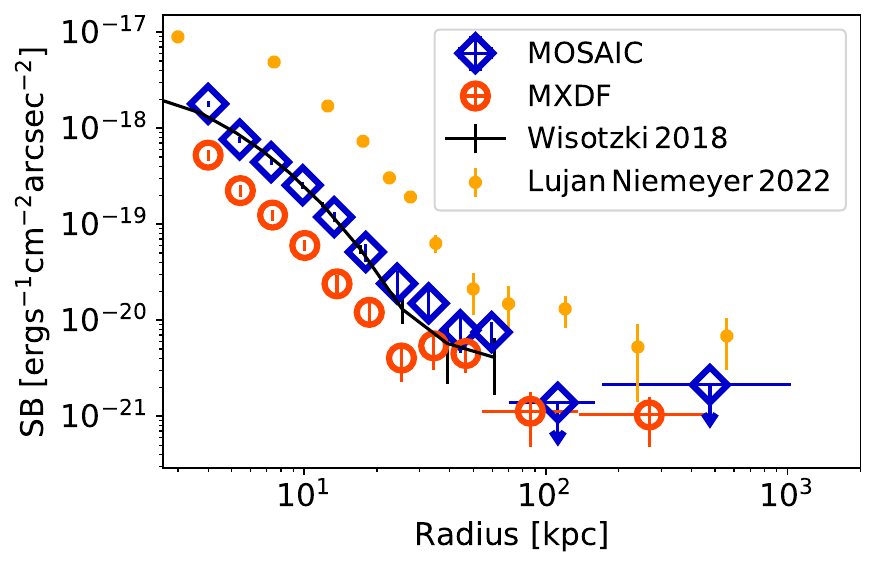}
\caption{\lya\ surface-brightness profiles of LAEs at $3<z<4$ in MOSAIC field (blue symbols) and MXDF field (red symbols). 
Both profiles include nearby LAEs.
For comparison, we also plot the result of \citet{wisotzki18} and \citet{maja22a}. The \citet{maja22a} profile is scaled by $(1+z)^4$.
\label{fig_profile_mosaic}}
\end{figure}

We have shown the median \lya\ surface brightness profiles of LAEs at $3<z<4$ out to $\approx$270~kpc using the MXDF data.
In this section, we firstly try to detect the \lya\ signal at larger distance.
The MOSAIC is a good supplement to the MXDF, with $\approx$9 times larger sky coverage, but with a 10-14 times shorter integration time.
We would like to use MOSAIC to constrain the median \lya\ surface brightness as far as 1000~kpc.
We present the radial profile in Figure~\ref{fig_profile_mosaic}.
We do not achieve a robust detection at the distance of 60-1000~kpc. Instead, we provide a two-sigma upper limit of $\mathrm{ 2.82 \times\ 10^{-21}\,erg\,s^{-1} \, cm^{-2} \, arcsec^{-2} }$.

In Figure~\ref{fig_profile_mosaic}, we compare with the \lya\ surface brightness profiles of \citet{wisotzki18}.
The profile of MOSAIC agrees well with \citet{wisotzki18} within approximately 60~kpc.
The median \lya\ profile of MXDF is about 0.5~dex fainter than that of the MOSAIC and \citet{wisotzki18}. 
Considering that the shallower observation of the MOSAIC can only identify the bright LAEs (Figure~\ref{fig_distributions}),
this comparison of \lya\ surface-brightness profiles in the MOSAIC and the MXDF again reflects the trend with $\mathrm{L_{Ly\alpha}}$ (see Figure~\ref{fig_bins}).

We also show the observational result of the Hobby-Eberly Telescope Dark Energy Experiment \citep[HETDEX,][]{gebhardt21}.
The HETDEX \lya\ surface-brightness profile around LAEs at $z\approx2.5$ is provided by \citet{maja22a} and shown by the yellow dots in Figure~\ref{fig_profile_mosaic}.
The \lya\ surface-brightness profile is scaled by $(1+z)^4$.
Our observation and HETDEX have different PSF sizes, spatial resolutions, and observational strategies. HETDEX performs shallow observations over a large sky area, while our MXDF and MOSAIC concentrate on small sky areas with extremely deep exposures. 
The LAE sample of \citet{maja22a} is one order of magnitude brighter than our sample. Their typical $\mathrm{L_{Ly\alpha}}$ is $\mathrm{10^{42.8} erg\,s^{-1}}$.
For comparison, the median $\mathrm{L_{Ly\alpha}}$ of the MXDF and MOSAIC samples at $3<z<4$ is $\mathrm{L_{Ly\alpha, MXDF} \approx 10^{41.1} erg\,s^{-1}}$ and $\mathrm{L_{Ly\alpha, MOSAIC} \approx 10^{41.5} erg\,s^{-1}}$, respectively.
In Figure~\ref{fig_profile_mosaic}, the profiles are strongly influenced by the different PSF.
\citet{maja22a} tried to re-scale the profile of \citet{wisotzki18} to their PSF size and LAE luminosity, and they obtained very good agreement.

Despite the different luminosities, observational facilities, and noise levels, the \lya\ surface-brightness profiles of the MXDF, MOSAIC, and \citet{maja22a} show similar patterns.
Among these datasets, the MXDF obviously achieves a deeper detection limit. Therefore, we took the radial profile of the MXDF as an example.
At small distances, the profile shows a power-law decrease.
The normalisation of the profile depends strongly on $\mathrm{L_{Ly\alpha}}$.
Then, the profile likely reaches a plateau.
The transition between the decrease and plateau appears at $\approx$20~kpc for MXDF.
The transition distance increases with higher $\mathrm{L_{Ly\alpha}}$.
Beyond the plateau ($\approx$50~kpc for MXDF), the profile likely drops to a very low surface-brightness level and becomes dominated by neighbouring galaxies (Section~\ref{subsec:neighbors}).
The distance of this break also increases with $\mathrm{L_{Ly\alpha}}$.

We note that a similar trend is also observed in the \lya\ surface-brightness profile of quasars at $2<z<4$, though the physical mechanisms behind this may be different. 
Several works find that the \lya\ surface-brightness profile within a distance of approximately 100~kpc of the quasar decreases by a power law with a power-law index of approximately -1.8 \citep[e.g.][]{borisova16,cai19}.
According to \citet{lin22}, the profile follows a similar power law within 1~cMpc. At larger radii (1-100~cMpc), the profile shows a flatter slope but with large variation, which can be well explained by a two-halo term of the clustered \lya\ sources. 

\section{Discussion} \label{sec:discussions}

\subsection{Physical origins of the LAHs}
\label{subsec:physical_origin}

While the existence of LAHs around star-forming galaxies at $2<z<6$ has been confirmed by previous works \citep[e.g.][]{steidel11,matsuda12,momose14,wisotzki16,wisotzki18,leclercq17,kusakabe22}, there is no consensus yet on the dominant physical mechanism giving rise to the LAHs.
Several physical origins are on the list \citep[e.g.][]{momose16,ouchi20}, including the resonant scattering of \lya\ photons produced by the recombination of gas photoionised by star formation or an AGN (in the central galaxy and also in unresolved satellites),
the collisional excitation and recombination in cooling gas flowing into a galaxy, 
as well as the fluorescent emission from the UV background.
The dominant origin of the \lya\ photons may change with the distance to the galaxies \citep[e.g.][]{lake15,mitchell20, byrohl21}.

In this work, we detected \lya\ emission around LAEs extending out to $\approx$270~kpc by median-stacking a representative sample of LAEs (Figure~\ref{fig_profile_all_unmask}), re-demonstrating that LAHs are a common property of LAEs.
Our measurement provides the typical \lya\ surface brightness profile around the LAE with median \lya\ luminosity $\mathrm{\approx 10^{41.1} erg\,s^{-1}}$.
In this section, we discuss the dominant physical origin of the \lya\ photons at different radial ranges.

Before discussing the physics dominating at different distances, we first clarify the difference between the extent of the LAHs and the size of dark-matter haloes.
Following the method of \citet{leclercq17}, we estimate the virial radius based on the $r_{vir}$ - UV magnitude relation predicted by the semi-analytic model of \citet{garel15}.
The typical virial radius $r_{vir}$ of the MXDF LAEs is approximately 20~kpc.
The estimation of $r_{vir}$ is always model-dependent, and we expect a large amount of scatter. 
Our detection of extended \lya\ emission out to $\approx$270~kpc implies that the \lya\ emission can extend to $\gtrsim $10 times larger radii than $r_{vir}$.

\subsubsection{$\mathrm{r \lesssim 20~kpc}$}
The production of \lya\ photons in the inner region of the LAH is closely related to the `central engine', e.g., the star formation in the galaxy.
The \lya\ photons are produced by recombination in the star-forming regions of the host galaxy; part of them succeed in escaping the dusty ISM. 
The existence of outflows may help the escape of \lya\ photons.
These escaped \lya\ photons can be scattered by the neutral hydrogen in the CGM and produce the observed LAH.
Spectroscopic observations find that the majority of \lya\ lines are redshifted with respect to the systemic redshift and have red-asymmetric profiles, which is interpreted as a signature of galactic outflows \citep[e.g.][]{verhamme06,schaerer11,dijkstra12,chang22}.
Such a model of \lya\ scattering through outflowing media has been successful in reproducing the observed \lya\ spectral and surface brightness profiles \citep[e.g.][]{yang16,song20,li22}.

In our analysis, within the central $\approx$20~kpc ($r_{vir}$), the \lya\ surface-brightness profile drops down as a power law with increasing radius (Figure~\ref{fig_profile_all_unmask}).
As is shown in Figure~\ref{fig_bins}, dividing the sample into high-$\mathrm{L_{Ly\alpha}}$ and low-$\mathrm{L_{Ly\alpha}}$ subsamples results in similarly shaped \lya\ surface-brightness profiles, but the normalisation increases with $\mathrm{L_{Ly\alpha}}$.
This is also seen in Figure~\ref{fig_profile_mosaic}, where we compare datasets spanning a larger range of $\mathrm{L_{Ly\alpha}}$.
If we assume that the \lya\ photons within this radius are primarily produced by star formation, the similar slopes of the high-$\mathrm{L_{Ly\alpha}}$ and low-$\mathrm{L_{Ly\alpha}}$ subsamples suggest that the production and propagation mechanism of the \lya\ photons in the inner CGM are similar for LAEs with different $\mathrm{L_{Ly\alpha}}$.

\subsubsection{$\mathrm{20 \lesssim r \lesssim 270~kpc}$}

{At a distance of $\mathrm{20 \lesssim r \lesssim 270~kpc}$, the \lya\ surface-brightness profile exhibits a flattening trend (Figure~\ref{fig_profile_all_unmask}), although we cannot completely rule out variations due to errors.}
The total \lya\ photon budget out to 270~kpc is too big to be explained solely by the \lya\ photons scattered from the central galaxy, as is shown in Section~\ref{subsec:neighbors}.
In Section~\ref{subsec:mosaic}, we also show that the shape of the \lya\ surface brightness profile appears similar for LAEs at different luminosities and redshifts, and even for quasars.
For MXDF LAEs, the flattening occurs at about 20~kpc, corresponding to 1~$r_{vir}$.
Based on absorption detected in the background quasar continuum, the KBSS survey \citep{rakic12} also found a similar radial evolution trend of the \lya\ optical depth for a sample of UV-selected star-forming galaxies at $z \approx 2.4$ with the stellar mass $\mathrm{M_* \approx 10^{10.1} M_\odot}$. 
The \lya\ optical depth (and hence the atomic hydrogen{ column density}) drops by more than an order of magnitude within 1~$r_{vir}$ ($\approx$100~kpc) and then stays enhanced at larger distances (out to at least 2.8~pMpc). 
Although based on different methods and different galaxy samples, \citet{rakic12} and our work seem to manifest a change of the state of the CGM at a distance of about 1~$r_{vir}$.
A detailed investigation of this change requires comparison with state-of-the-art simulations \citep[e.g.][Guo et al. in preparation]{blaizot23} as well as better observations. 
Below we present our speculative considerations concerning the underlying physics responsible for the \lya\ photon budget at $\mathrm{20 \lesssim r \lesssim 270~kpc}$.

At 20-50~kpc ($\approx$1-3~$r_{vir}$), the \lya\ emission is  robustly detected.
It is difficult to simply attribute the flattening trend to the contribution of (bright) neighbours.
At this radial span, the \lya\ surface-brightness profile shows only marginal inconsistency after masking the nearby LAEs detected within the field of view (Figure~\ref{fig_profile_compare_mask}). As is shown in Figure~\ref{fig_LAE_neighbors}, neighbouring LAEs start to emerge at tens of kiloparsecs, but the majority are at hundreds of kiloparsecs. 
The interpretation of the \lya\ surface-brightness profile at this distance (20-50~kpc) is therefore complicated, as multiple factors are involved: the neighbours start to have an effect, and the `central engine' starts to lose its dominance.
The flattening of the \lya\ profile probably indicates a change in the dominant power source, which can be explained as signs of cooling radiation or undetected satellites, or a mixture of all \citep[e.g.][]{haiman00,rosdahl12,mitchell20}.
The \lya\ emission can also be explained as the fluorescence emission from gas illuminated by the UV background \citep[e.g.][]{cantalupo05,ribas16}.
For example, \citet{gallego21} used the \lya\ emission at large distance to constrain the photoionisation rate of hydrogen and the covering fraction of Lyman limit systems, based on the assumption that all the \lya\ emission at large distances originates from \lya\ fluorescence in optically thick \hi\ clouds.

Compared to 20-50~kpc, the \lya\ surface-brightness profile is likely to drop at approximately 50~kpc, and then it stays at a lower surface-brightness level out to 270~kpc (Figure~\ref{fig_profile_all_unmask}).
This potential break at approximately 50~kpc is clearer after we remove the nearby LAEs. 
In Figure~\ref{fig_profile_compare_mask}, after masking the nearby LAEs within the same NB, the \lya\ emission at 50-270~kpc is below the detection limit.
The comparison of the two profiles in Figure~\ref{fig_profile_compare_mask} thus shows that bright, nearby LAHs make a major contribution to the observed \lya\ surface-brightness profile at 50-270~kpc ($\gtrsim 3r_{vir}$).

In the simulations of \citet{mitchell20}, the LAHs at $3<z<4$ at tens of kiloparsecs are mainly powered by satellite galaxies.
The neighbours defined in this work (Section~\ref{subsec:neighbors}) are not necessarily satellites, they are namely galaxies bounded within the gravitational potential of a more massive central  galaxy.
The clustering analysis of \citet{herrero23} suggests that only $\lesssim$10\%\ of the MXDF LAEs (same as our LAE sample) are satellites.
Therefore, in our case, the neighbours that contribute to the LAHs at 50-270~kpc are mostly bright central galaxies instead of satellites.
\citet{byrohl21} studied a similar case.
They applied Monte Carlo radiative transfer of \lya\ to TNG50 cosmological magnetohydrodynamical simulations.
Their simulations find that most \lya\ photons at large radii originate from nearby bright haloes and are then scattered in the CGM of the target halo and also in the IGM.
In our work, taking advantage of deep exposures and high spatial resolution, we directly observe the LAHs being enhanced by nearby bright LAEs (Figure~\ref{fig_bins}).

Therefore, at radii of 50-270~kpc, we can confirm that the star formation in nearby bright LAEs provides a major contribution to the extended \lya\ emission.
The contribution of other \lya\ emission mechanisms at these radii, including cooling radiation, flourescence, and satellite galaxies, cannot be robustly measured.
Instead, we provide an upper limit on the total of these mechanisms (Section~\ref{subsec:neighbors}).

The potential break observed in the \lya\ surface-brightness profile at approximately 50~kpc is intriguing.
Limited by the S/N, we cannot rule out the possibility of error variations. Here, we provide a few speculative considerations. 
It seems to indicate a quick change in the powering mechanism of the LAH or physical state of the CGM at this distance.
This may suggests a typical clustering pattern of satellites or enhancement of the \lya\ flux due to collisional excitation of gas inflows at $\mathrm{20 \lesssim r \lesssim 50~kpc}$. Another possibility is that the neutral hydrogen density decreases at approximately 50~kpc. 
According to the simulations of \citet{mitchell20}, the hydrogen number density shows a hint of decrease with radius, but their simulations stop at 1~$r_{vir}$.
The gas state and physical processes at a large distance from the galaxy (around 3~$r_{vir}$ in our case) are not very clear. 
To further address this question, we conducted detailed simulations that will be described in Guo et al. in preparation.

\subsection{The \lya\ surface-brightness profile as a cross-correlation function}
\label{subsec:correlation}

Intensity mapping of the cumulative \lya\ and other emission lines (e.g. \hi, CO) has been proposed as a a powerful tool for understanding the line emissivity and the large-scale matter distribution in the Universe \citep[e.g.][]{bernal22}.
In particular, by stacking the cross-correlation functions (CCFs) between the source redshift catalogue and maps expected to include \lya\ emission from the same sources, efforts have been made to constrain the integrated \lya\ emission over large cosmological volumes \citep[e.g.][]{croft16,croft18,kakuma22,kikuchi22,lin22}.
Based on the Sloan Digital Sky Survey (DR12) spectra, \citet{croft16,croft18} measured the cross-correlation between quasar positions and \lya\ emission imprinted in the residual spectra of luminous red galaxies.
They detected positive signal of extended \lya\ emission around quasars on scales as far as 15~cMpc.
This measurement was renewed by \citet{lin22} using DR16 of the Sloan Digital Sky Survey.
The extended \lya\ emission around normal galaxies is orders of magnitude fainter than the quasars and thus more difficult to detect.
Recently, positive detections of the \lya-LAE CCF as far as $\approx$1~cMpc have been obtained by stacking deep NB images around bright LAEs \citep{kakuma22,kikuchi22}.

In the two-dimensional case, the CCF between \lya\ emission intensity and the position of a given LAE mathematically equals the \lya\ surface brightness at different radii \citep[e.g.][]{kakuma22}.
Restricted by the sky area of the MXDF and MOSAIC, our observations are not ideal for mapping the extended emission at a scale of several comoving megaparsecs,
but the extremely deep exposure and good spatial resolution provide a unique experiment that identifies very faint galaxies and foreground contaminators.
This could be a good supplement to \lya\ intensity mapping experiments.
Our work is more efficient in removing line and continuum contaminators. Our dedicated pseudo-NBs include all the potential signal but not extra noise.
More importantly, we quantify the contribution of the faint LAEs that were not detected in previous works and anonymously contribute to their \lya\ CCFs.

Our precise and reliable measurement of the \lya\ surface brightness profile will be helpful for the design of future intensity mapping experiments, which usually have lower spatial resolution and perform much shallower observations.
It would also be very useful to compare our CCFs with cosmological simulations.

\section{Summary} \label{sec:summary}
Thanks to the extremely deep MUSE observations of the Hubble Ultra Deep Field, 
we unveil the typical \lya\ surface-brightness profile around LAEs down to an unprecedented depth and distance.
Our major results are summarised below.

Based on the MXDF data, we present the median \lya\ surface-brightness profiles of LAEs with \lya\ luminosity $\mathrm{\approx 10^{41.1} erg\,s^{-1}}$ at $3<z<4$. After carefully correcting for the systematic surface-brightness offsets of the MUSE data cube, we detect extended \lya\ emission out to 270~kpc. 
The \lya\ surface-brightness profile decreases as a power law within a radius of 20~kpc, followed by a flattened profile at 20-50~kpc and a potential drop to a lower level at 50-270~kpc.
We observe a possible break in the \lya\ profile at a radius of approximately 50~kpc.

We find that the nearby LAEs make a major contribution to the \lya\ surface-brightness profile at 50-270~kpc, at the level of $\mathrm{ 1.20 \pm 0.49 \times\ 10^{-21}\,erg\,s^{-1} \, cm^{-2} \, arcsec^{-2} }$. In addition, we provide an upper limit on the contribution of other physical mechanisms to the profile at this distance, which is estimated to be $\mathrm{ 0.93 \times\ 10^{-21}\,erg\,s^{-1} \, cm^{-2} \, arcsec^{-2} }$.

We divide our LAE sample in subsamples by $\mathrm{L_{Ly\alpha}}$. 
Within approximately 20~kpc, the high-$\mathrm{L_{Ly\alpha}}$ sample exhibits a higher \lya\ surface-brightness profile, but the profiles of different subsamples have similar slopes.
This indicates that star formation in the central galaxy may dominate the \lya\ surface brightness at a small distance. The production and propagation of \lya\ photons likely follow similar mechanisms in different subsamples.

We attempted to detect the extended \lya\ emission at larger radii using MOSAIC.
We are not able to provide a robust detection from 60~kpc to 1~Mpc, but we provide an upper limit of $\mathrm{ 2.82 \times\ 10^{-21}\,erg\,s^{-1} \, cm^{-2} \, arcsec^{-2} }$.

Although this work mainly focuses on the LAHs at $3<z<4$, we find that there is no significant evolution in the observed \lya\ surface brightness profiles when we compare with $4<z<5$ and $5<z<6$.

Our results support a scenario in which star formation in the central galaxy dominates the LAHs at small radii (within 20~kpc), while \lya\ photons from nearby galaxies dominate the \lya\ surface brightness at large radii (50-270~kpc).
The \lya\ surface-brightness profile at the distance range of 20-50~kpc is more difficult to interpret. Deeper observations of the \lya\ line profiles and further simulations are needed.

\begin{acknowledgements}
Y.G., R.B. and L.W. acknowledge support from the ANR/DFG grant L-INTENSE (ANR-20-CE92-0015, DFG Wi 1369/31-1).
LW acknowledges support by the ERC Advanced Grant SPECMAG-CGM (GA101020943).

\end{acknowledgements}

\bibliographystyle{aa} 
\bibliography{main}

\end{document}